\pdfoutput=1
\documentclass[pre,twocolumn,amsmath,amssymb,aps]{revtex4}
\usepackage{graphicx}
\usepackage{dcolumn}
\usepackage{bm}
\usepackage{textcomp}
\usepackage{color,xcolor}

\begin{document}
\title{Nernst-like Effect in a Flexible Chain}

\author{Shuji Tamaki}
\author{Keiji Saito}

\affiliation{
	Department of Physics, 
	Keio University, 
	Yokohama 223-8522, 
	Japan}

\date{\today}

\begin{abstract}
We investigate heat transport via a charged flexible chain in the presence of magnetic fields.
We focus on the Nernst-like effect, where the average positions of particles deviate in the perpendicular direction to the heat flow.
This phenomenon is induced by the nonlinear dynamics as well as nonequilibrium state.
We develop a linear response formalism to derive a thermodynamic force which induces the Nernst-like effect, and show that the phenomenon is quantitatively explained. 
We also discuss the inverse effect, where an external ac-driving force induces finite net heat current in the homogeneous system attached to heat baths with the same temperature.
\end{abstract}

\maketitle

\section{\label{sec:intro}introduction}

Heat transport in a mesoscopic scale has been intensively studied experimentally and theoretically. Heat transport via a small system has exhibited a number of intriguing phenomena, such as Kondo effect \cite{saito2013kondo}, conductance quantum \cite{rego1998quantized,schwab2000measurement,chiu2005ballistic}, and anomalous heat transport \cite{lepri-livi-politi-2003,dhar-2008,lepri-2016,lepri-livi-politi-1997,lepri-livi-politi-1998,casati-prosen-2003,basile-olla-bernardin-2006,mai-dhar-narayan-2007,saito-dhar-2010,vanbeijeren-2012,spohn-2014,shen-henry-tong-zheng-chen-2010,wang-carter-lagutchev-koh-seong-cahill-dlott-2007,meier-menges-nirmalraj-holscher-riel-gotsmann-2014,chang-okawa-garcia-majumdar-zettl-2008,lee-wu-lou-lee-chang-2017}, to name only a few. Recent technological development has also opened the field of controlling heat in small systems \cite{li-ren-wang-zhang-hanggi-li-2012,casati2005}.

We also note that magnetic fields can cause many nontrivial effects on the heat transfer. One of such effects includes the thermal Hall effect \cite{goldsmid-2010,strohm-rikken-wyder,sheng-sheng-ting,kagan-maksimov,inyushkin,lifa-1,lifa-2,lifa-3,katsura,onose,ideue}, which shows an emergence of transverse heat current to applied thermal gradient in the presence of magnetic fields. This effect has many variations depending on heat carriers, such as, the phonon Hall effect in paramagnetic dielectrics \cite{strohm-rikken-wyder,sheng-sheng-ting,kagan-maksimov,inyushkin,lifa-1,lifa-2,lifa-3}, the magnon Hall effect in ferromagnetic dielectrics \cite{katsura,onose} and multiferroics \cite{ideue}. Another example related to the magnetic field is the Nernst effect \cite{goldsmid-2010,stark-brandner-saito-seifert-2014}, where the magnetic field induces transverse electric voltage under heat flow in the electric systems.

In this paper, we consider several effects induced by magnetic fields in small charged flexible chains, which have never been addressed seriously so far. We note that typical low-dimensional objects such as nanofibers \cite{shen-henry-tong-zheng-chen-2010}, polymers \cite{wang-carter-lagutchev-koh-seong-cahill-dlott-2007, meier-menges-nirmalraj-holscher-riel-gotsmann-2014}, and Carbon-nanotubes \cite{chang-okawa-garcia-majumdar-zettl-2008,lee-wu-lou-lee-chang-2017} possess finite charges. In general, the magnetic fields bend the direction of motion via the Lorentz force, and hence one may anticipate some positional effect at the macroscopic level. In equilibrium case, the positional distribution is described by the canonical distribution and one can show that no particular effect on particle's position appear. However, as we discuss in this paper, the nonequilibrium situation with nonlinear forces drastically change the positional distribution, and finite deviation in the perpendicular direction to the applied thermal gradient appears.
We call this {\em Nernst-like effect in a flexible chain} (NEFC) from the analogy to the Nernst effect in electric systems \cite{goldsmid-2010,stark-brandner-saito-seifert-2014}.
We present similarities and dissimilarities between the NEFC and the original Nernst effect in the electronic systems.
It is shown that this effect is induced by nonlinear dynamics. This implies that a simple linearized model such as the Rouse model \cite{deGennes} is not available to discuss this phenomenon. The effective force that causes the NEFC is a thermodynamic force. We provide a systematic analysis to obtain the thermodynamic force within the linear response theory. The theory quantitatively reproduces the deviations of the particles.

We also consider an inverse phenomenon under finite magnetic fields, i.e., a generation of a finite heat current in equilibrium situation by inducing the transverse deviation in the flexible chain. We apply an oscillating fields to induce finite positional deviation in the transverse direction on average. 

This paper is organized as follows. In Sec.~\ref{sec:model}, we introduce a model. In Sec.~\ref{sec:temp}, we describe the numerical methods and show typical temperature profile at the steady state. In Sec.~\ref{sec:position}, we discuss the NEFC in the charged flexible chain. In the section \ref{sec:linear-response}, we develop a formalism to obtain the effective thermodynamic force causing the NEFC with the linear response theory. In Sec.~\ref{sec:inverse}, we demonstrate the inverse effect of the NEFC. Finally, we summarize our study in Sec.~\ref{sec:discussion}.

\begin{figure}[t]
\includegraphics[width=8.6cm]{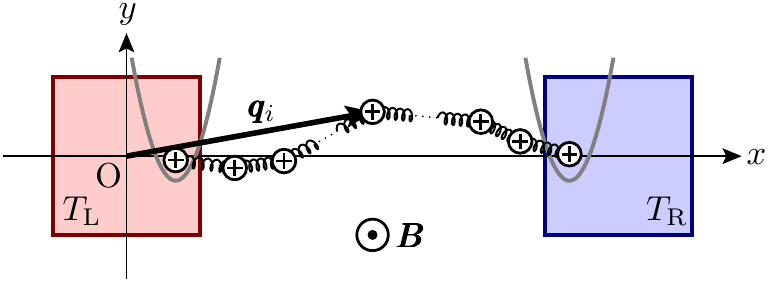}
\caption{\label{fig:model}Schematic of the system connected to left and right heat baths at temperatures $T_{\text{L}}$ and $T_{\text{R}}$. The system can move on the $xy$-plane and a constant magnetic field $\bm{B}$ is applied in the $z$-direction. End particles are trapped by a pinning potential $\phi(r)$ represented by the gray curves. The bold black arrow represents the position vector $\bm{q}_{i}$ of the $i$th particle.}
\end{figure}

\section{\label{sec:model}Setup}

\subsection{\label{subsec:setup}Model}

We consider a flexible chain connected to two heat baths at its ends. The schematic of the model is shown in Fig. \ref{fig:model}.
The system is composed of $N$ particles with mass $m$, and neighboring particles are connected by spring forces.
For simplicity, we assume that the motion of the particles is confined on the $xy$-plane.
The position and canonical momentum of the $i$th particle are denoted by $\bm{q}_{i}=(q_{i,x},q_{i,y})$ and $\bm{p}_{i}=(p_{i,x},p_{i,y})$, respectively.
We consider the case where all the particles are uniformly charged with the charge $e$ and a static magnetic field $B$ is applied in the $z$-direction.
The Hamiltonian of the system is described as follows
\begin{eqnarray}
H
&=&
\sum_{i=1}^{N}\frac{|\bm{p}_{i}-e\bm{A}(\bm{q}_{i})|^{2}}{2m}
+
\sum_{i=1}^{N-1}V(|\bm{r}_{i+1,i}|)\nonumber\\
&+&
\frac{e^{2}}{4\pi\varepsilon_{0}}\sum_{i=2}^{N}\sum_{j=1}^{i-1}\frac{1}{|\bm{r}_{i,j}|}
+
\sum_{i=0\,\text{and}\,N}\phi(|\bm{r}_{i+1,i}|),
\label{eq:hamiltonian}
\end{eqnarray}
where $\bm{r}_{i,j}:=\bm{q}_{i}-\bm{q}_{j}$ is the stretch vector, which is a relative vector between the $i$th and $j$th particles.
The vector $\bm{A}=(A_{x},A_{y})$ is the gauge potential.
One of the simplest form of the gauge potential is ${\bm A}({\bm q}_i)= -{\bm q}_i \times {\bm B}/2$, although the detailed choice of the gauge does not affect physical results as long as it satisfies $\partial A_{y}/\partial q_{x}-\partial A_{x}/\partial q_{y}=B$. The actual velocity is expressed as $\bm{v}_{i}:=\dot{\bm{q}}_{i}=\left(\bm{p}_{i}-e\bm{A}(\bm{q}_{i})\right)/m$.

The terms $V(|{\bm r}|)$ and $\phi (|{\bm r}|)$ respectively stand for the spring potential between neighbor sites and the pinning potential to bind end particles.
We assume the following simple forms for these potentials:
\begin{eqnarray}
V(|{\bm r}|)
=
\frac{k}{2}(|{\bm r}|-\ell)^{2},
\quad\text{and}\quad
\phi(|{\bm r}|)
=
\frac{k}{2} |{\bm r}|^{2} \, .
\label{eq:spring-potential}
\end{eqnarray}
The parameter $\ell$ is the natural length of the spring.
We impose the fixed boundary condition, i.e., we fix the positions of the end particles as $\bm{q}_{0}=(\ell,0)$ and $\bm{q}_{N+1}=(N\ell,0)$, which respectively bind ${\bm q}_1$ and ${\bm q}_N$ around the same positions due to the potential $\phi(|{\bm r}|)$. Hence, in the equilibrium situation, the particles are located along the $x$-direction on average.
The third term in the Hamiltonian stands for the Coulomb interaction.

The left and right reservoirs are respectively attached to the first and the $N$th particles, and those have the temperature $T_{\rm L}$ and $T_{\rm R}$.
We model the dynamics of heat reservoirs by the Langevin thermostat \cite{allen-tildesley-1987}. Then, the total dynamics is given by the deterministic dynamics from the Hamiltonian (\ref{eq:hamiltonian}) and the Langevin dynamics for the end particles as follows
\begin{eqnarray}
m\dot{v}_{i,x}
&=&
F_{i+1,i,x}-F_{i,i-1,x}
+
\frac{e^{2}}{4\pi\varepsilon_{0}}
\sum_{\substack{j=1\\j\neq i}}^{N}
\frac{r_{i,j,x}}{|\bm{r}_{i,j}|^{3}}
+
eBv_{i,y}\nonumber\\
&+&
\delta_{i,1}\left(-\gamma v_{i,x}+\eta_{\text{L},x}\right)
+
\delta_{i,N}\left(-\gamma v_{i,x}+\eta_{\text{R},x}\right),
\label{eq:EOM-x}\\
m\dot{v}_{i,y}
&=&
F_{i+1,i,y}-F_{i,i-1,y}
+
\frac{e^{2}}{4\pi\varepsilon_{0}}\sum_{\substack{j=1\\j\neq i}}^{N}\frac{r_{i,j,y}}{|\bm{r}_{i,j}|^{3}}
-
eBv_{i,x}\nonumber\\
&+&
\delta_{i,1}\left(-\gamma v_{i,y}+\eta_{\text{L},y}\right)
+
\delta_{i,N}\left(-\gamma v_{i,y}+\eta_{\text{R},y}\right).
\label{eq:EOM-y}
\end{eqnarray}
Here, $r_{i,j,\alpha}$ is the $\alpha$th component in the vector ${\bm r}_{i,j}$ $(\alpha=x,y)$. The vector $\bm{F}_{i+1,i}=(F_{i+1,i,x},F_{i+1,i,y})$ is the spring force vector given by
\begin{align}
\bm{F}_{i+1,i}
&:=
\begin{cases}
k\bm{r}_{i+1,i},&\text{for}\quad i=0,N\vspace{3mm}\\
k(|\bm{r}_{i+1,i}|-\ell)\dfrac{\bm{r}_{i+1,i}}{|\bm{r}_{i+1,i}|},&\text{for}\quad i=1,...,N-1\\
\end{cases}. \nonumber\\
\label{eq:spring-pinning-forces}
\end{align}
The variable $\gamma$ in Eqs.(\ref{eq:EOM-x}) and (\ref{eq:EOM-y}) is a friction constant. The random variables $\bm{\eta}_{\mu}=(\eta_{\mu,x},\eta_{\mu,y})~(\mu={\rm L,R})$ are the Gaussian white noises satisfying the fluctuation dissipation theorem
\begin{eqnarray}
\bm{\langle}\!\langle
\eta_{\mu,\alpha}(t)\eta_{\mu ',\alpha '}(s)
\rangle\!\bm{\rangle}
=
2\gamma k_{\rm B}T_{\mu}\delta_{\mu\, \mu'}\delta_{\alpha\, \alpha'}\delta(t-s) \, , 
\end{eqnarray}
where $\bm{\langle}\!\langle ... \rangle\!\bm{\rangle}$ denotes the noise average.

\subsection{\label{subsec:remark}Remark on the potentials and a parameter set}

We here remark on the potentials (\ref{eq:spring-potential}). Although the potential form of $V(|{\bm r}|)$ and $\phi (|{\bm r}|)$ is quadratic in terms of $|{\bm r}|$, these potentials produce a nonlinear force in terms of positional variables ${\bm q}_i$, as long as the natural length $\ell$ is finite.
Note that as explicitly written in (\ref{eq:spring-pinning-forces}), the relative distance $|{\bm r}_{i+1,i}|$ in the force $F_{i+1,i,\alpha} \, (\alpha=x,y)$
causes a nonlinear force in terms of the position variables $\bm{q}_{i}$ for finite $\ell$.
Throughout this paper, we fix the parameters $(m,k,\ell)=(1,1,10)$ unless newly defined, and consider several variations of $e B$ and $e^{2}/4\pi\varepsilon_{0}$. The friction coefficient is set to $\gamma=1$. The Boltzmann constant $k_{\rm B}$ is set to $1$ in the numerical calculation.
In the nonequilibrium simulation, we typically use the temperature set $(T_{\rm L}, T_{\rm R})=(2,1)$ \cite{tltr}.

\section{\label{sec:temp}Temperature profile in the presence of magnetic fields}

In this section, we briefly show a steady state temperature profile in the presence of magnetic fields. We define two types of local temperatures by the kinetic energies of the $x$ and $y$ component of velocities, i.e.,
\begin{align}
T_{i,\alpha}&:={ m\langle v_{i,\alpha}^{2}\rangle  }\, ~~~ (\alpha=x,y).
\end{align}
To get the average values of kinetic energies, we first check that the system achieves the steady state from a uniformity of the local heat current along the chain, and next compute a long time average. In general, when a magnetic field is applied to the system, the dynamics becomes much more complicated than the dynamics without magnetic fields. Hence, careful computation is necessary by checking numerical errors. Details of numerical method, sensitivity of numerical error on magnetic fields, and check of achievement of the steady state are shown in Appendix \ref{ap:sc:simulation}.
We use the modified velocity Verlet algorithm~\cite{allen-tildesley-1987} with a finite time discretization $\varDelta t=10^{-3}$.
For the system size $N=256$, $10^{8}$ time steps are necessary to reach the steady state and about $10^{8}$ time step is required for the long-time average.

\begin{figure}[t]
\includegraphics[width=8.6cm]{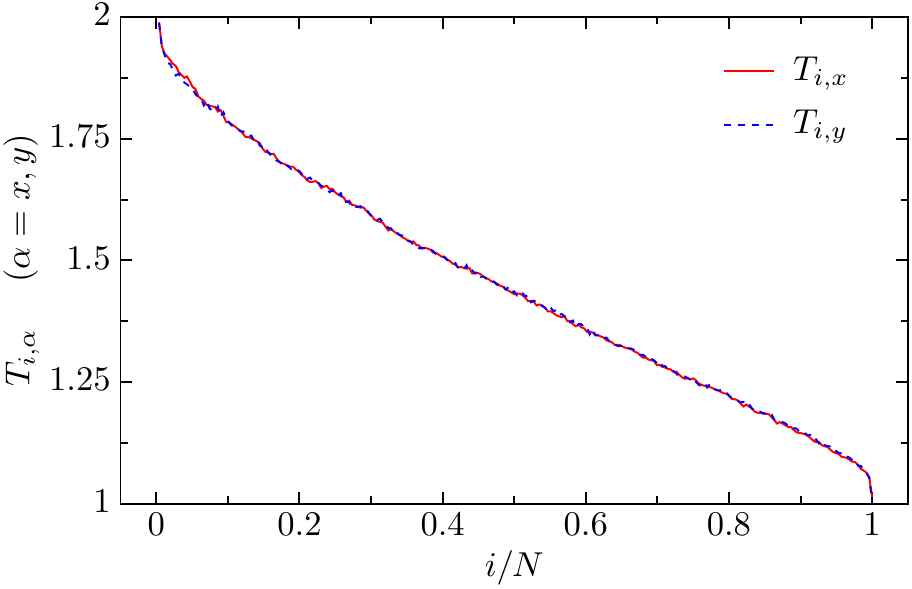}
\caption{\label{fig:TB}The local temperature profile for the system size $N=256$. The parameters are set to $(T_{\rm L},T_{\rm R})=(2,1)$, $e^{2}/4\pi\varepsilon_{0}=10$, and $eB=0.5$.}
\end{figure}

In Fig. \ref{fig:TB}, we present the local temperature profiles, $T_{i,x}$ and $T_{i,y}$ for the case of $eB=0.5$. The figure shows that the two temperature profiles agree with each other. Hence, a local temperature is uniquely determined. The temperature profile is smooth except to the boundaries. Although the present system size is not very large to obtain an asymptotic behavior of the temperature profile, the uniqueness of local temperatures are expected to hold even in the relatively small system sizes.

\section{\label{sec:position}Average position in the presence of magnetic fields}
In this section, we investigate an average position of each particle in the nonequilibrium steady state.

\subsection{\label{subsec:equilibrium}Equilibrium case}
We first discuss the equilibrium situation setting $T_{\rm L}=T_{\rm R}=T$. In this case, the stationary distribution is expressed by the canonical distribution:
\begin{eqnarray}
P_{\rm eq}(\bm{\varGamma})
&=&
\frac{ e^{-H( \bm{\varGamma } )/T }  }{Z}
\quad\text{with}\quad
Z:=
\int d\bm{\varGamma}\,e^{- H(\bm{\varGamma})/T}, ~~
\label{eq:canonical-distribution}
\end{eqnarray}
where $\bm{\varGamma}$ stands for the phase space variables composed of the positions and velocities of all the particles, i.e., $\bm{\varGamma}:=(\{\bm{q}_{i}\}_{i=1}^{N},\{\bm{v}_{i}\}_{i=1}^{N})$. The distribution is invariant under the transformation $\{ q_{i,y} \} \leftrightarrow \{ -q_{i,y} \}$. Hence the average position in the $y$-direction is zero, i.e., 
\begin{eqnarray}
\langle q_{i,y}\rangle
=
\int d\bm{\varGamma}\,q_{i,y}P_{\text{eq}}(\bm{\varGamma})
=
0.
\end{eqnarray}
Hence, the particles are aligned on the $x$-axis on average.

\subsection{\label{subsec:position}Nonequilibrium case}

\begin{figure}[t]
\includegraphics[width=8.6cm]{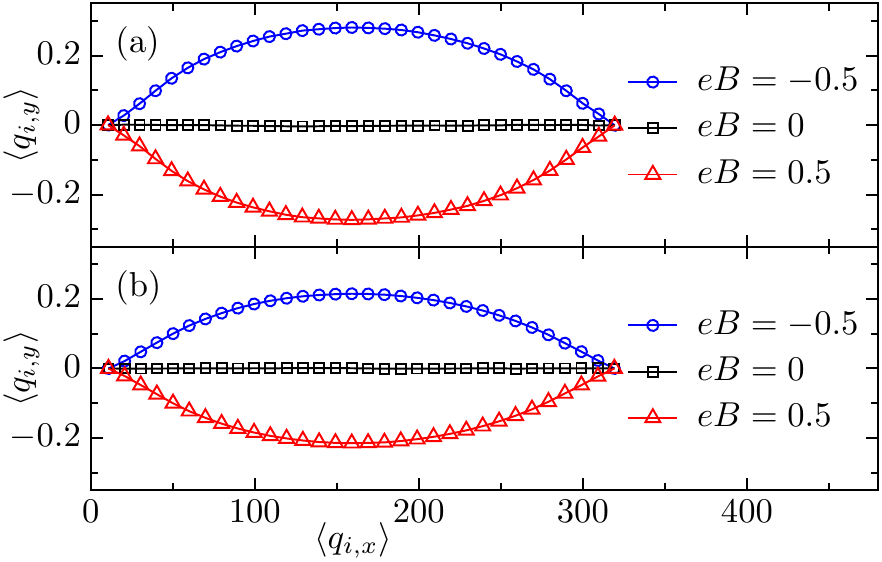}
\caption{\label{fig:nernst}
 Long-time average of the positional configuration for the different magnetic fields $eB=0$ (black squares), $eB=0.5$ (red triangles), and $eB=-0.5$ (blue circles). The system size is $N=32$ and the temperature set is $(T_{\text{L}},T_{\text{R}})=(2,1)$. Figures (a) and (b) show the case of $e^{2}/4\pi\varepsilon_{0}=10$ and $e^{2}/4\pi\varepsilon_{0}=0$, respectively.}
\end{figure}

We next discuss the nonequilibrium situation using the numerical calculation. The method of numerical simulation is the same as in Sec. \ref{sec:temp}.
In Fig.~\ref{fig:nernst}(a), we show the average position of each particle at the steady state for the system $e^{2}/4\pi\varepsilon_{0}=10$ with the size $N=32$.
In the case of zero magnetic field, the particles are perfectly aligned on the $x$-axis. However, when a finite magnetic field is applied, a finite deviation in the $y$-direction appears. If the magnetic field is reversed, the direction of deviation is also reversed. Note that this finite deviation in the $y$-direction cannot occur in the equilibrium situation as discussed in Sec. \ref{subsec:equilibrium}. Hence this phenomenon is clearly a {\em nonequilibrium effect}. 

Although we have considered the simple and small system in Fig.~\ref{fig:nernst}(a), one can check that qualitatively the same phenomena appear in more general cases, such as the case where the particles can move in the three-dimensional space (not confined on the $xy$-plane), the case of larger system sizes with various magnetic fields, and other temperature sets. We present these demonstrations in the Appendices \ref{app:3D}-\ref{app:deltat}.

Below, we discuss whether a simplified dynamics can explain the phenomenon and thereby investigate its mechanical origin. We first consider the role of the Coulomb interaction. Let us switch off the Coulomb interaction, i.e., $e^{2}/4\pi\varepsilon_{0}=0$ while keeping $eB=\pm 0.5$. We show the result for this case in Fig.~\ref{fig:nernst}(b). The result is qualitatively the same as the result in Fig.~\ref{fig:nernst}(a). Hence, the finite deviation in the $y$-direction can occur even without the Coulomb interaction.

Next, we consider more simplified model. We note that in polymer physics, a linear dynamics such as the Rouse model \cite{deGennes} successfully explains many dynamical aspects.
Motivated by this, let us employ the linearized model and discuss whether it can generate the finite deviation in the $y$ direction. We dare to set $\ell =0$ in our Hamiltonian, where the dynamics becomes completely linear. (Note that as remarked in Sec. \ref{subsec:remark}, a finite natural length induces a nonlinear spring force between the particles.) Suppose that we expand the system by a finite length $L$ in the $x$-direction, i.e., we set the boundary condition ${\bm q}_0 = (0,0)$ and ${\bm q}_{N+1} = (L,0)$.
Then, let us consider the long time average of the position of each particle, solving the following equations of motion
\begin{eqnarray}
m\dot{v}_{i,x}&=&k(q_{i+1,x}+q_{i-1,x}-2q_{i,x})+eBv_{i,y}\nonumber\\
&+&\delta_{i,1}(-\gamma v_{i,x}+\eta_{\text{L},x})
+\delta_{i,N}(-\gamma v_{i,x}+\eta_{\text{R},x}),~~~~
\label{eq:EOM-x-harmonic}\\
m\dot{v}_{i,y}&=&k(q_{i+1,y}+q_{i-1,y}-2q_{i,y})-eBv_{i,x}\nonumber\\
&+&\delta_{i,1}(-\gamma v_{i,y}+\eta_{\text{L},y})
+\delta_{i,N}(-\gamma v_{i,y}+\eta_{\text{R},y}).~~~~
\label{eq:EOM-y-harmonic}
\end{eqnarray}
Here, note that trivial long time averages $\langle \dot{\bm{v}}_{i}\rangle=\langle\bm{v}_{i}\rangle=\bm{0}$ and $\langle \eta_{\mu,\alpha} \rangle = 0$ hold for any temperature set $(T_{\text{L}},T_{\text{R}})$. Using these properties, one can immediately obtain a long-time average $\langle q_{i,\alpha} \rangle$ from a straightforward calculation. 
The results are given as $\langle q_{i,x}\rangle=iL/(N+1)$ and $\langle q_{i,y}\rangle=0$, regardless of the temperature set and amplitude of the magnetic field. From this argument, we form a conclusion that the magnetic field does not affect the average configuration as long as the dynamics contains only linear forces. From this consideration, we claim that a \textit{nonlinear dynamics} is necessary to obtain the finite deviation observed in Fig~\ref{fig:nernst}.
This also indicates that it seems to be difficult to make a simple theoretical model to explain the phenomenon, since it requires real nonlinear dynamics. 

The finite deviation observed in the charged flexible chain is reminiscent of the Nernst effect in the electronic systems~\cite{goldsmid-2010}.
The Nernst effect is observed when a magnetic field is applied to an electronic system with a finite heat flow. The effect is detected as 
the emergence of a finite bias voltage in the perpendicular direction to the heat flow.
This bias voltage is very similar to the transverse deviation observed in the present case.
However, there are several differences between the original Nernst effect in an electronic system and the present phenomenon.
The original Nernst effect can appear even without any interaction between the particles~\cite{stark-brandner-saito-seifert-2014}.
However, the present case requires many-body interaction.
In addition, a nonlinear force is necessary for the present phenomenon.
Hence, to discriminate the present effect from the original Nernst effect, we call it \textit{the Nernst-like effect in a flexible chain} (NEFC).

\section{\label{sec:linear-response}Thermodynamic effective force}
\subsection{\label{sec:formalism}Linear response formalism}

The finite deviation in the transverse direction means the existence of the force in the $y$-direction. 
We here analyze this effective force in the NEFC. This effective force should originate from the Lorentz force, since the magnetic field enters only in this term.
However, we should note that the effective force is not determined simply by the long-time average of the Lorentz force, as it is always zero, i.e.,  $\langle\bm{v}_{i}\times\bm{B}\rangle=\langle\bm{v}_{i}\rangle\times\bm{B}=\bm{0}$.
Rather, we consider this with the thermodynamic argument.
To this end, we conduct a linear response analysis for small temperature difference between heat baths.

Let $\mathcal{F}_{i}$ be a thermodynamic effective force for the $i$th particle in the presence of the magnetic field.
To obtain $\mathcal{F}_{i}$, we apply an additional external force $f_{i}$ to the $i$th particle such that the average deviation $\langle q_{i,y}\rangle$ vanishes.
From the force balance, we can identify the effective force by the relation $\mathcal{F}_{i}=-f_{i}$.
We formulate this framework in the linear response regime. Let $j_{i+1,i}$ be a local heat current between the $i$th and $(i+1)$th particles. See Appendix \ref{ap:sc:simulation} for the explicit expression of the local heat current $j_{i+1,i}$. When the temperature difference $\varDelta T=T_{\text{L}}-T_{\text{R}}$ and external forces $f_{i}$ are small, the average heat current $\langle J \rangle:=(N-1)^{-1}\sum_{i=1}^{N-1}\langle j_{i+1,i}\rangle$ and the average transverse deviation $\langle q_{i,y} \rangle $ can be expressed as follows
\begin{eqnarray}
\begin{bmatrix}
\langle J\rangle\\
\langle q_{1,y}\rangle\\
\vdots\\
\langle q_{N,y}\rangle
\end{bmatrix}
=\begin{bmatrix}
L_{00}&L_{01}&\cdots&L_{0N}\\
L_{10}&L_{11}&\cdots&L_{1N}\\
\vdots&\vdots&\ddots&\vdots\\
L_{N0}&L_{N1}&\cdots&L_{NN}
\end{bmatrix}
\begin{bmatrix}
\varDelta T\\
f_{1}\\
\vdots\\
f_{N}
\end{bmatrix}.
\label{eq:Onsager-expansion}
\end{eqnarray}
Here, the matrix $\bm{L}$ is a $(N+1)\times(N+1)$ linear-response matrix. Once the matrix ${\bm L}$ is obtained, the column vector $[\varDelta T, f_{1},\cdots,f_{N}]^{T}$ is given as a function of $\langle J\rangle$ and $\langle q_{i,y}\rangle$ by inverting the matrix $\bm{L}$, i.e., $[\varDelta T, f_{1},\cdots,f_{N}]^{T} = {\bm L}^{-1} [\langle J\rangle, \langle q_{1,y}\rangle, \cdots , \langle q_{N,y}\rangle ]^T$. Finally, we set $\langle q_{i,y}\rangle=0$ to obtain the thermodynamic force $\mathcal{F}_i$ through the force balance $\mathcal{F}_{i}=-f_{i}$. 

Each element of the linear-response matrix is given by the following expression
\begin{alignat}{2}
L_{00}(B)&=\frac{1}{ T^{2}}\int_{0}^{\infty} dt\,\langle J(t)J(0)\rangle_{\text{eq}}^{B},\quad&
\label{eq:L00}\\
L_{i0}(B)&=\frac{1}{ T^{2}}\int_{0}^{\infty} dt\,\langle q_{i,y}(t)J(0)\rangle_{\text{eq}}^{B},\quad& i\neq 0,
\label{eq:Li0}\\
L_{0j}&=0,\quad& j\neq 0,
\label{eq:L0i}\\
L_{ij}&=\frac{1}{ T}\langle q_{i,y}q_{j,y}\rangle_{\text{eq}},\quad& i,j\neq0.
\label{eq:Lij}
\end{alignat}
For the derivation of each expression, see Appendix~\ref{app:linear-response}.
Here, the symbol $\langle ... \rangle_{\text{eq}}$ represents the equilibrium average with respect to the canonical distribution~(\ref{eq:canonical-distribution}). Note that the time-evolution in the correlation functions includes the contribution of heat baths, i.e., the time-evolution is given by solving the equations of motion (\ref{eq:EOM-x}) and (\ref{eq:EOM-y}) with an equal temperature $T_{\rm L}=T_{\rm R}=T$.
The superscript $B$ in Eqs.~(\ref{eq:L00}) and~(\ref{eq:Li0}) represents a finite magnetic field $B$.
The element $L_{00}$ satisfies the Onsager-Casimir symmetry: $L_{00}(B)=L_{00}(-B)$.
We comment that the element $L_{ij}$ ($i,j\neq 0$) is independent of magnetic field, because the potential part in the equilibrium distribution is independent of $B$.

From Eq.~(\ref{eq:L0i}), the heat current is expressed as $\langle J\rangle=L_{00}\varDelta T$.
Then, setting $\langle q_{i,y}\rangle=0$, the thermodynamic force is written as
\begin{eqnarray}
\mathcal{F}_{i} = - f_i &=&
-(\bm{L}^{-1})_{i0}\langle J\rangle\nonumber\\
&=&
-(\bm{L}^{-1})_{i0}L_{00}\varDelta T.
\label{formulaF}
\end{eqnarray}
When we do not apply the external force $f_i$, the deviation in the $y$-direction induced by the finite temperature is given by the 
formula
\begin{eqnarray}
\langle q_{i,y} \rangle &= L_{i0} \Delta T \, . \label{qiy}
\end{eqnarray}

\begin{figure}[t]
\includegraphics[width=8.6cm]{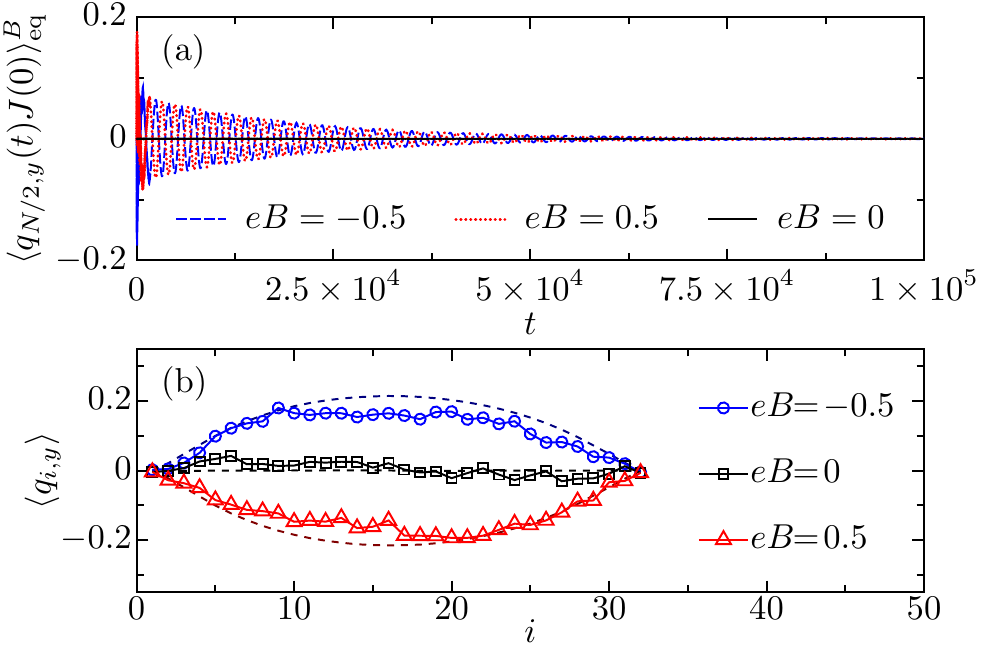}
\caption{\label{fig:lr}Demonstration of the formula (\ref{eq:Li0}). We set $e^{2}/4\pi\varepsilon_{0}=0$. (a) Equilibrium time-correlation between deviation of the central particle ($i=N/2$) and heat current. The temperature is set to $T_{\text{L}}=T_{\text{R}}=1.5$. (b) Average transverse deviation $\langle q_{i,y}\rangle=L_{i0}\varDelta T$ calculated using Eq.~(\ref{eq:Li0}). The dashed lines indicate the values calculated by performing nonequilibrium simulation with the temperature set to $(T_{\text{L}},T_{\text{R}})=(2,1)$; these data are identical to those in Fig.~\ref{fig:nernst}(b).}
\end{figure}

\subsection{\label{sec:effective}Calculation of the effective force}

We here check the validity of the linear response formalism and investigate the properties on the thermodynamic effective force.
We first demonstrate that the linear response formula~(\ref{eq:Li0}) reproduces the results by the direct nonequilibrium simulation in Sec. \ref{sec:position}.
For simplicity of the numerical computation, we consider the case of zero Coulomb interaction, i.e., $e^{2}/4\pi\varepsilon_{0}=0$.
In Fig.~\ref{fig:lr}(a), we show $\langle q_{N/2,y}(t)J(0)\rangle_{\text{eq}}^{B}$ as a typical example of the equilibrium time-correlation function.
The figure shows that the time-correlation is almost zero for the case of zero magnetic field, while the finite magnetic fields generate a damped oscillation.
The linear response values $\langle q_{i,y}\rangle$ are calculated from the formula (\ref{qiy}) and those are shown in Fig.~\ref{fig:lr}(b) (symbols and solid lines).
We also plot the results by the nonequilibrium simulation in the same figure (dashed lines). 
Note that the results by the nonequilibrium simulation are identical to Fig. \ref{fig:nernst}(b). 
One can see that the linear response result displays quantitatively adequate agreement with that of the nonequilibrium simulation.
This agreement also guarantees that the present temperature set $((T_{\rm L},T_{\rm R})=(2,1))$ is within the range of the linear response. See also the Appendix~\ref{app:deltat} on the temperature dependence of the NEFC.

\begin{figure}[t]
\includegraphics[width=8.6cm]{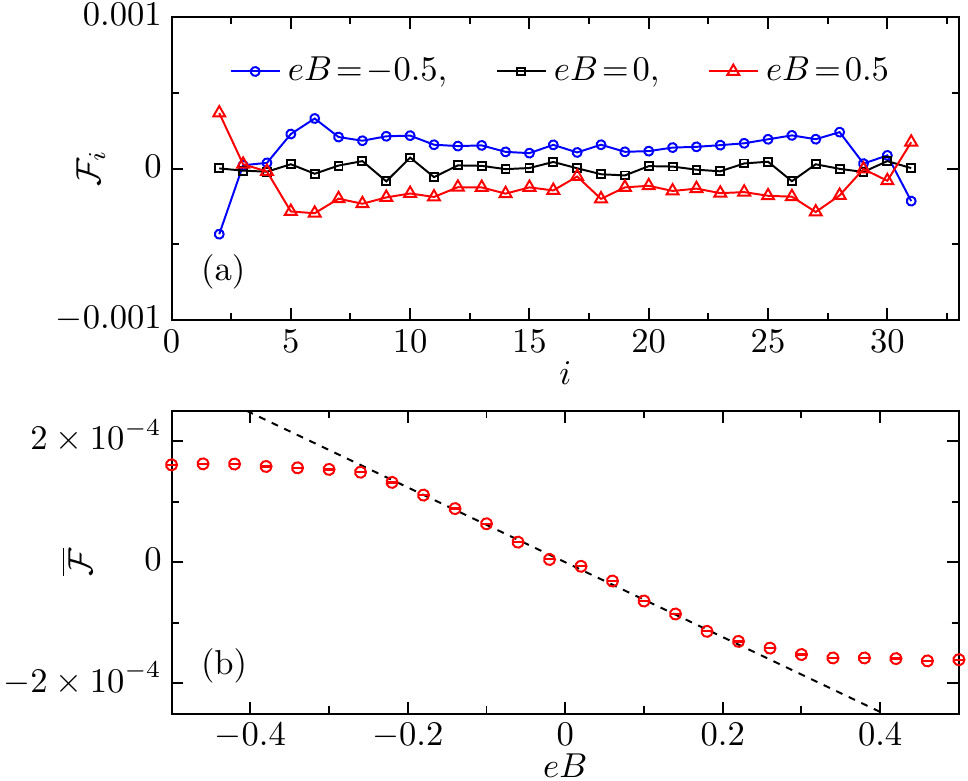}
\caption{\label{fig:f}Effective forces for the parameter set in Fig.~\ref{fig:nernst}(b). Figure (a) shows the results for the magnetic fields $eB=0$ (black squares and line), $eB=0.5$ (red triangles and line), and $eB=-0.5$ (blue circles and line). Figure (b) shows the magnetic-field dependence of the effective force averaged over bulk particles except for five particles each from both edges, i.e., $\overline{\mathcal{F}}:=\sum_{i=6}^{N-5}\mathcal{F}_{i}/(N-10)$. The dashed line is a guiding line to highlight the linear behavior $\overline{\mathcal{F}}\propto B$.}
\end{figure}

Having obtained the reliability of the linear response calculation, we now show the thermodynamic effective forces for the finite magnetic fields with the formula (\ref{formulaF}). We present the numerical results for the magnetic fields $eB=0$, and $\pm0.5$ in Fig.~\ref{fig:f}(a). In the case of zero magnetic field, the effective force is almost zero. On the other hand, finite magnetic fields generate finite effective forces. When the magnetic field is reversed, the effective forces are also reversed. It is also observed that the effective forces are almost uniform except for the edges.

We note that the steady state temperature profile shows a finite temperature gradient as shown in Fig. \ref{fig:TB}, which is roughly uniform except to the boundaries. Within the linear response regime, one may expect that the local effective force depends linearly on the local temperature gradient. This expectation seems to be consistent with that the local effective force in the bulk is almost uniform. In addition, we show the magnetic-field dependence of the effective force averaged over bulk particles in Fig. \ref{fig:f}(b). The effective force exhibits linear dependence for small magnetic fields. Combining these arguments, the effective force is likely to behave as $\mathcal{F}_{i}\propto B(\nabla T)_{i}$, as long as the magnetic field is small.

\subsection{\label{sec:phenomenological}Phenomenological argument on the profile of average positions}

We finally form a phenomenological theory based on the above numerical observations. To this end, we regard that our system behaves as a {\em string} with a constant effective tension $t_{\text{eff}}$. Then, for small deviations in the transverse direction, we can use the effective potential in the continuous picture~\cite{fetter-walecka-1980}:
\begin{eqnarray}
U_{\text{eff}}\sim\int_{0}^{L}d\xi\,
\left[
\frac{t_{\text{eff}}}{2}\left(\frac{\partial q_{y}(\xi)}{\partial\xi}\right)^{2}+\mathcal{F} q_{y}(\xi)
\right],
\end{eqnarray}
where $\xi=i\ell$ and $L=N\ell$. The function $q_{y}(\xi)$ is the transverse deviation at the position $\xi$. We consider the fixed boundary condition, i.e., $q_{y}(0)=q_{y}(L)=0$. The variable $\mathcal{F}$ is the effective force. From the numerical observations, we assumed that $\mathcal{F}$ is independent of the position $\xi$ for small temperature gradient. Minimizing $U_{\text{eff}}$ with respect to $q_{y}$, we obtain the optimized profile:
\begin{eqnarray}
q_{y}(\xi)\sim\frac{\mathcal{F}}{2t_{\text{eff}}}\xi(\xi-L).
\label{phenomenological}
\end{eqnarray}
The overall structure of this profile is a parabola and hence is consistent with the numerical observation in Figs. \ref{fig:nernst} and \ref{fig:lr}.

\section{\label{sec:inverse}Inverse effect}

In this section, we consider the inverse effect of the NEFC. That is, we demonstrate that a finite net heat current along the chain can be generated in the presence of magnetic fields by making finite deviation of particles in the transverse direction even in the equilibrium situation.

\subsection{\label{subsec:setup}Setup}

\begin{figure}
\includegraphics[width=8.6cm]{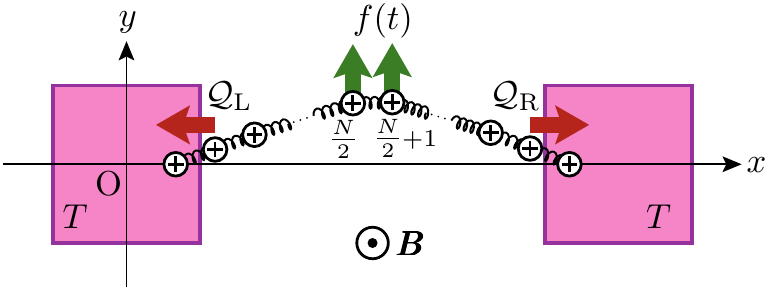}
\caption{\label{fig:inverse} Schematic of the setup for the demonstration of inverse effect. We set the temperatures of two heat baths to an equal value $T$ and apply an ac-driving force $f(t)$ only on two central particles, i.e., $i=N/2$, $N/2+1$. The heat current is measured at the contact between the system and the heat baths.}
\end{figure}

The setup for the demonstration of the inverse effect is schematically depicted in Fig. \ref{fig:inverse}.
We set an equal temperature for the two heat baths. We apply an external driving force in the $y$-direction on the two central particles $i=N/2$ and $i=N/2+1$ for simplicity ($N$ is an even integer). Hence equations of motion are given by Eqs.(\ref{eq:EOM-x}) and (\ref{eq:EOM-y}) with additional driving force terms only for these two particles.
Since we consider the equilibrium situation setting equal temperature in the baths, we need to inject a finite work into the system in order to obtain a finite net heat current from one heat bath to another. Obviously, if the external force is static in time, a finite heat current cannot be generated because no work can be injected into the system. Hence, we consider the following ac-driving force:
\begin{eqnarray}
f(t)=f_{0}\sin^{2}\left(\frac{2\pi}{\tau}t\right).
\label{eq:external-force}
\end{eqnarray}
Note that the force $f(t)$ has the same sign for any time. Hence this force leads to finite deviations of particles in the $y$-direction on average. Then the question is whether this type of driving force in the presence of magnetic fields can generate a finite net heat current from one heat bath to another.
Note that in the absence of magnetic field, one can show that the long-time average of the net heat current is zero from the reflection symmetry in the spatial structure~\cite{fn-2}.

We measure the heat current at the contact between the system and the heat baths, i.e., 
\begin{eqnarray}
j_{\rm L}
&:=&
\bm{v}_{1}\cdot(-\gamma\bm{v}_{1}+\bm{\eta}_{\rm L}),
\label{eq:jL}\\
j_{\rm R}
&:=&
\bm{v}_{N}\cdot(-\gamma\bm{v}_{N}+\bm{\eta}_{\rm R}).
\label{eq:jR}
\end{eqnarray}
Then, we focus on the heat flowing from the system into the heat baths during the period $\tau$,
\begin{eqnarray}
\mathcal{Q}_{\mu}(\tau):=-\int_{0}^{\tau} dt\,j_{\mu}(t)
\quad\text{with}\quad
\mu=\text{L, R}.
\label{eq:def-Q}
\end{eqnarray}
The difference between these quantities $\mathcal{Q}_{\text{L}}-\mathcal{Q}_{\text{R}}$ is the net heat transferred from the right to the left heat bath.

\subsection{\label{subsec:B-tau}Numerical results on the net heat current}

\begin{figure}[t]
\includegraphics[width=8.6cm]{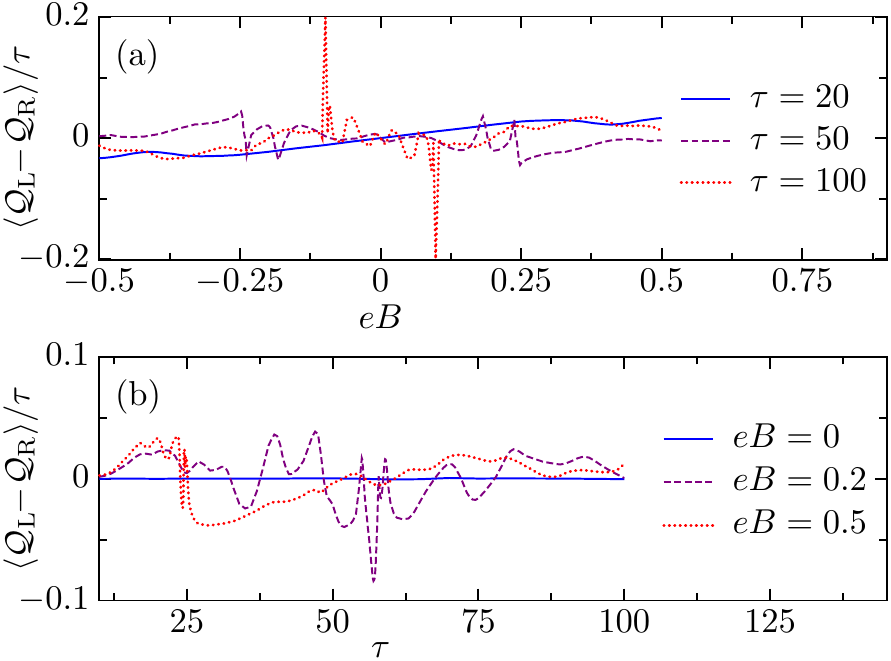}
\caption{\label{fig:QBt}Plot of net heat current for the system size $N=32$. The parameters are $e^{2}/4\pi\varepsilon_{0}=0$ and $T_{\text{L}}=T_{\text{R}}=1.5$. Figure (a) shows the magnetic-field dependence for the periods $\tau=20$, 50, and 100. Figure (b) shows the period-dependence for the magnetic fields $eB=0$, 0.2, and 0.5. Error bars are within the line width.}
\end{figure}

We consider the system $e^{2}/4\pi\varepsilon_{0}=0$ with the size $N=32$. The temperatures of the two heat baths are set to $T_{\text{L}}=T_{\text{R}}=1.5$. The amplitude of the external force is set to $f_{0}=1$. 

We first show the magnetic-field dependence of the net heat current in Fig.~\ref{fig:QBt}(a) for the periods $\tau=20$, 50, and 100. A finite net current is observed for finite magnetic fields. For a small-period case, the average net current is small and depends linearly on the magnetic fields when the magnetic fields are small. As the period $\tau$ increases, the behavior of the net current becomes complicated as a function of $B$. The sign of the net current also can change. The net current satisfies the symmetry; it reverses its sign with the magnetic-field reversal.

Next, we consider the period dependence of the net heat current for the magnetic fields $eB=0$, 0.2, and 0.5 in Fig.~\ref{fig:QBt}(b). In case of zero magnetic field, the net heat current is zero for any values of $\tau$. Whereas, for finite magnetic fields, a finite net current is observed. Particularly, for the small period region, the dependence on the magnetic fields is not systematic.

Finally, we comment on the role of nonlinear forces in the inverse effect. The inverse effect requires the nonlinear forces as in the NEFC. We can prove that the inverse effect can not appear for the case of linear forces as shown in Appendix~\ref{a:sec:harmonic-chain}.

\section{\label{sec:discussion}Summary and discussion}

In this study, we investigated the heat transport phenomena in a charged flexible chain in the presence of a constant magnetic field.
Main findings include the NEFC and the inverse effect. We emphasize that a nonlinear force is a key ingredient for these effects.
We also develop a linear response formalism to obtain the thermodynamic force inducing the NEFC.

We here comment on the role of nonlinear force to get the NEFC. In the section \ref{sec:position}, we proved that the linear dynamics cannot show the NEFC. We note that the case of $\ell=0$ is the complete harmonic chain. As well known \cite{rieder-lebowitz-lieb-1967}, the harmonic chain cannot form a finite temperature gradient. This indicates that the system does not possess the local equilibration in the nonequilibrium state. Taking account of this fact, one may think that the NEFC requires the local equilibration process in the dynamics. However, we have one counterexample that cannot show the NEFC even when the equilibration process exists. That is a toy model recently proposed in Refs. \cite{tamaki-sasada-saito-2017,saito-sasada-2018}. This model has no nonlinear forces in the spring potential. Rather, it contains so-called conserving noises to induce the local equilibration, keeping conserved quantities. This model shows a finite temperature gradient in the nonequilibrium situation. Nevertheless, it is easy to show that the NEFC can not occur in this model, since the equations of motion contain only linear variables. The absence of the NEFC is proven by following the similar procedure done for linear equations (\ref{eq:EOM-x-harmonic}) and (\ref{eq:EOM-y-harmonic}). This implies that local equilibration is a necessary albeit insufficient condition for the NEFC. The NEFC requires real nonlinear dynamics.

The aim of this study is to provide a theoretical idea rather than to provide experimental implementation. Main aim is to show the existence of the NEFC and the inverse effect.
At present, it is not convenient to estimate an experimentally accessible setup, because it is not straightforward to obtain realistic values of spring constant, etc.
However, we hope that recent technological development enables us to observe the NEFC in realistic materials such as DNA molecule \cite{smith-finzi-bustamante-1992} with strong magnetic fields \cite{miura-herlach-1985} in future.

\begin{acknowledgments}
The authors would like to thank Taro Hanazato for performing some of numerical computations. They also thank Kunimasa Miyazaki for continuous interest in our study, and Yoshihiro Murayama for useful comments on experimental setup. We were supported by JSPS Grants-in-Aid for Scientific Research (No. JP25103003, JP16H02211 and JP17K05587).
\end{acknowledgments}

\appendix

\section{\label{ap:sc:simulation}DETAILS OF NUMERICAL SIMULATION}

\begin{figure}[t]
\includegraphics[width=8.6cm]{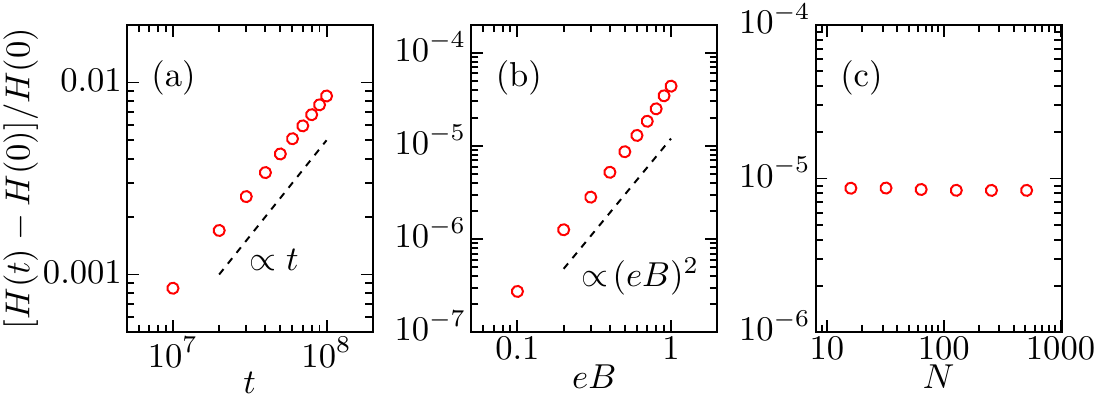}
\caption{\label{fig:error}Relative error of total energy in an isolated system owing to the finite time step $\varDelta t=10^{-3}$. (a): $t$-dependence for $N=32$ and $eB=0.5$, (b): magnetic-field dependence for $N=32$ and $t=10^{5}$, (c): system-size dependence for $eB=0.5$ and $t=10^{5}$.}
\end{figure}

Numerical simulation is performed by the modified velocity Verlet algorithm~\cite{allen-tildesley-1987} with the time discretization $\varDelta t=10^{-3}$.
In general, a magnetic field enhances complexity in the dynamics. 
We estimate the degree of numerical error by examining the conservation of the total energy for the Hamiltonian (\ref{eq:hamiltonian}) in an isolated system. We found that the numerical error ($[H(t)-H(0)]/H(0)$) increases in proportion to time $t$ as shown in Fig.~\ref{fig:error}(a), when the magnetic field is applied. It is also observed that the error is proportional to $B^{2}$ as shown in Fig.~\ref{fig:error}(b). However, it does not significantly depends on the system size as shown in Fig.~\ref{fig:error}(c). For the time $t=10^{8}$, $[H(t)-H(0)]/H(0)\approx0.0085$.

\begin{figure}[t]
\includegraphics[width=8.6cm]{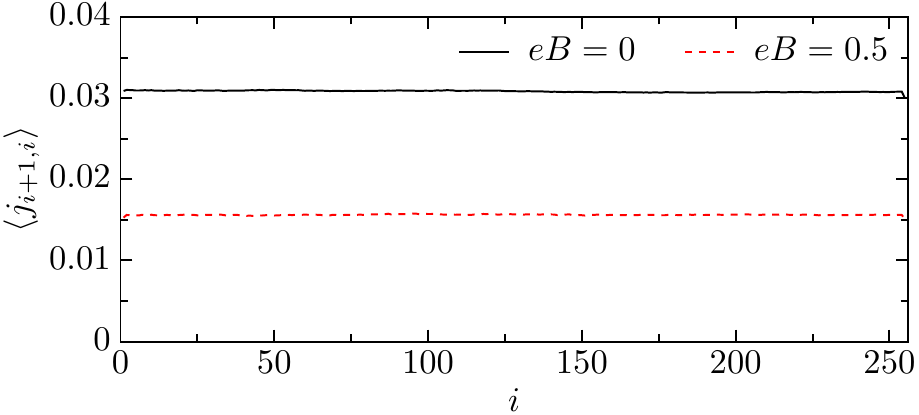}
\caption{\label{fig:j} Plot of heat current $j_{i+1,i}$ for the system size $N=256$. The parameter to control the strength of the Coulomb interaction is set to $e^{2}/4\pi\varepsilon_{0}=10$.}
\end{figure}

To demonstrate that the steady state is achieved for the open system setup, we check a uniformity of the local heat current. We define a local energy at the site $i$ as
\begin{eqnarray}
\varepsilon_{i}
:=
\frac{m|\bm{v}_{i}|^{2}}{2}
+
V(|\bm{r}_{i+1,i}|)
+
\sum_{\substack{j=1\\j\neq i}}^{N}
\frac{e^{2}}{4\pi\varepsilon_{0}}
\frac{1}{|\bm{r}_{i,j}|}.
\end{eqnarray}
Then, the heat current is defined through the continuity equation 
\begin{eqnarray}
j_{i+1,i}
&=&
-\bm{F}_{i+1,i}\cdot\bm{v}_{i+1}
\nonumber\\
&+&
\sum_{j=1}^{i}\sum_{j'=i+1}^{N}
\frac{e^{2}}{4\pi\varepsilon_{0}}
\frac{\bm{r}_{j,j'}}{|\bm{r}_{j,j'}|^{3}}
\cdot
\frac{\bm{v}_{j}+\bm{v}_{j'}}{2} \, .
\end{eqnarray}
This expression satisfies $\dot{\varepsilon}_{i}=j_{i,i-1}-j_{i+1,i}$. After approximately $10^{8}$ time steps, we compute the local heat currents. The result for the system size $N=256$ is shown in Fig.~\ref{fig:j}. The figure clearly shows a uniformity of the local heat current, which indicates an achievement of steady state in both zero and finite magnetic field.

\section{\label{app:3D}THE NEFC IN THREE-DIMENSIONAL SPACE}

In the main text, the motion of each particle is confined on the $xy$-plane. Here, we demonstrate that the NEFC also occurs when each particle moves in the three-dimensional space. We consider the system composed of $N$ particles. Let $\bm{q}_{i}=(q_{i,x},q_{i,y},q_{i,z})$ and $\bm{v}_{i}=(v_{i,x},v_{i,y},v_{i,z})$ respectively be a position and velocity vector. The equations of motion are written as
\begin{eqnarray}
m\dot{v}_{i,x}&\!=\!&
F_{i+1,i,x}-F_{i,i-1,x}
+\frac{e^{2}}{4\pi\varepsilon_{0}}\sum_{\substack{j=1\\j\neq i}}^{N}\frac{r_{i,j,x}}{|\bm{r}_{i,j}|^{3}}
+eBv_{i,y}\nonumber\\
&+&\delta_{i,1}\left(-\gamma v_{i,x}+\eta_{\text{L},x}\right)
+\delta_{i,N}\left(-\gamma v_{i,x}+\eta_{\text{R},x}\right),
\label{ap:eq:EOM-x}~~~~~~\\
m\dot{v}_{i,y}&\!=\!&
F_{i+1,i,y}-F_{i,i-1,y}
+\frac{e^{2}}{4\pi\varepsilon_{0}}\sum_{\substack{j=1\\j\neq i}}^{N}\frac{r_{i,j,y}}{|\bm{r}_{i,j}|^{3}}
-eBv_{i,x}\nonumber\\
&+&\delta_{i,1}\left(-\gamma v_{i,y}+\eta_{\text{L},y}\right)
+\delta_{i,N}\left(-\gamma v_{i,y}+\eta_{\text{R},y}\right).
\label{ap:eq:EOM-y}~~~~~~\\
m\dot{v}_{i,z}&\!=\!&
F_{i+1,i,z}-F_{i,i-1,z}
+\frac{e^{2}}{4\pi\varepsilon_{0}}\sum_{\substack{j=1\\j\neq i}}^{N}\frac{r_{i,j,z}}{|\bm{r}_{i,j}|^{3}}
\nonumber\\
&+&\delta_{i,1}\left(-\gamma v_{i,z}+\eta_{\text{L},z}\right)
+\delta_{i,N}\left(-\gamma v_{i,z}+\eta_{\text{R},z}\right).
\label{ap:eq:EOM-z}~~~~~~
\end{eqnarray}
We set the parameters to the same values as in the main text, i.e., $e^{2}/4\pi\varepsilon_{0}=10$, $(T_{\rm L},T_{\rm R})=(2,1)$ and $\gamma=1$.

\begin{figure}[b]
\includegraphics[width=8.6cm]{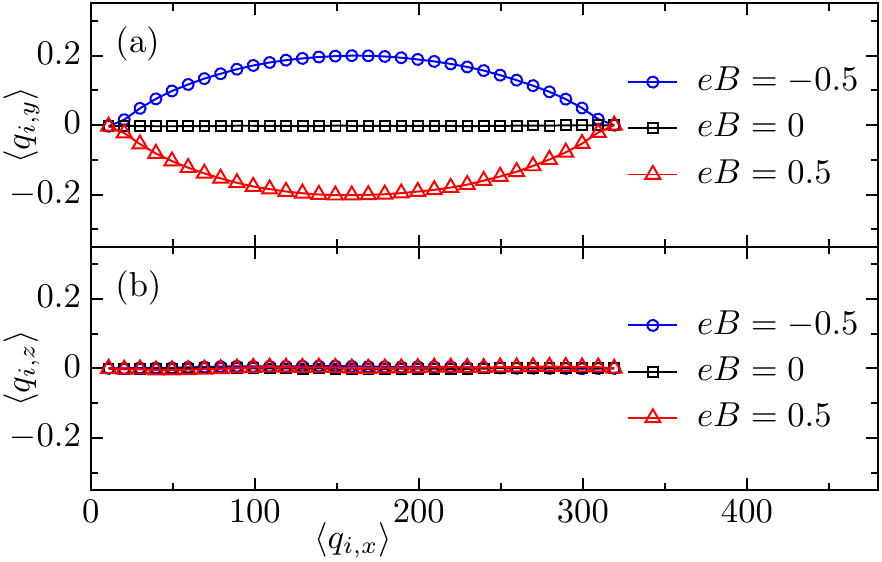}
\caption{\label{fig:3D-nernst}
Long-time average of the positional configuration for the magnetic fields $eB=0$ (black squares), $eB=0.5$ (red triangles), and $eB=-0.5$ (blue circles). Figure(a) and (b) shows $(\langle q_{i,x}\rangle,\langle q_{i,y}\rangle)$ and $(\langle q_{i,x}\rangle,\langle q_{i,z}\rangle)$, respectively.}
\end{figure}

We show average position of the particles in Fig.~\ref{fig:3D-nernst}. Figure~\ref{fig:3D-nernst}(a) shows the finite deviation in the $y$-direction. While, as shown in Fig.~\ref{fig:3D-nernst}(b), the $z$-component of the position is not affected by magnetic fields. Thus, the deviation is perpendicular to both the chain direction and the magnetic field.

\begin{figure}[b]
\includegraphics[width=8.6cm]{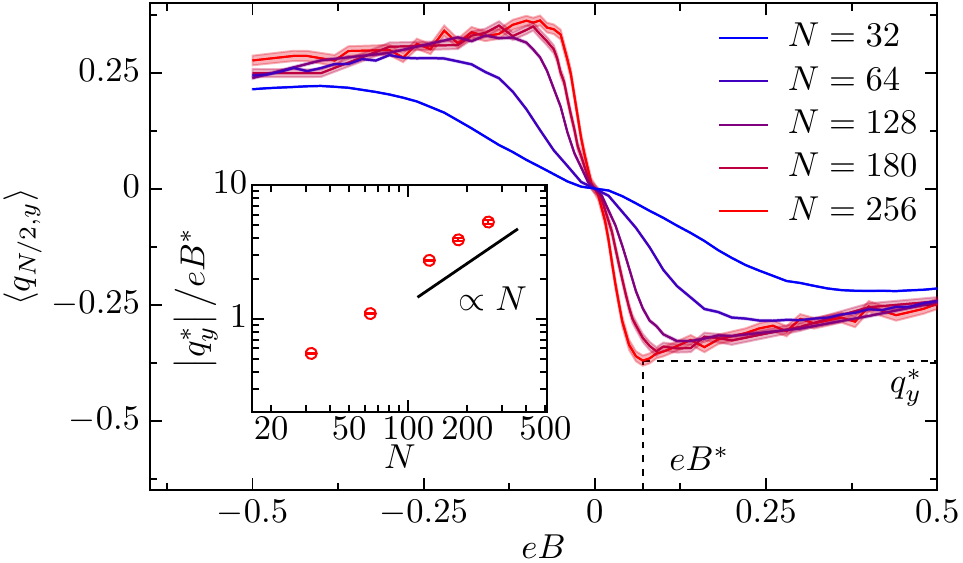}
\caption{\label{fig:yB}Magnetic-field dependence of transverse deviation $\langle q_{N/2,y}\rangle$ for different system sizes $N=32$, 64, 128, 180, and 256. The shaded region indicates the statistical error. For $N=256$, the minimum value of $\langle q_{N/2,y}\rangle$ is denoted by $q_{y}^{*}$, and the magnetic field at that point is $B^{*}$. The inset shows the system-size dependence of $|q_{y}^{*}|/eB^{*}$, which demonstrates the linear dependence on the system size $N$.}
\end{figure}

\section{\label{app:m-N-dependence}Magnetic-field and system-size dependence in NEFC}

We discuss the magnetic-field dependence of the transverse deviation in the NEFC, and also discuss its system-size dependence. 
Here, we consider the simplest case described by the Hamiltonian (\ref{eq:hamiltonian}) with $e^{2}/4\pi\varepsilon_{0}=0$.

Figure \ref{fig:yB} shows the magnetic-field dependence of the deviation of the central particle $\langle q_{N/2,y}\rangle$ for different sizes $N=32, 64, 128, 180$, and $256$. Temperatures are set to $(T_L,T_R)=(2,1)$. We can see the symmetry $\langle q_{N/2,y}\rangle\leftrightarrow -\langle q_{N/2,y}\rangle$ for the magnetic field reversal $B\leftrightarrow -B$. For a fixed system size, the overall structure of $\langle q_{N/2,y}\rangle$ is nonmonotonic with respect to the magnetic field, while they are monotonic for sufficiently magnetic field.

Let us consider the system-size dependence of the NEFC for small magnetic fields looking at $\langle q_{N/2,y}\rangle$. For small magnetic fields, we note that the amplitude of the transverse deviation roughly shows linear dependence on the magnetic field. Keeping this in mind, we define
\begin{align}
q_{y}^{*} &:= \min_{B} \langle q_{N/2,y}\rangle\, , \\
B^{*} &:= {\rm argmin}  \langle q_{N/2,y}\rangle \, .
\end{align}
We numerically calculate these quantities for each system-size. See the example for $N=256$ indicated in the figure. We calculate the quantity $|q_{y}^{*}|/B^{*}$ as a function of the system-size. This quantity roughly provides the system-size dependence of the amount of transverse deviation for a fixed small magnetic field. The result is shown in the inset of Fig.~\ref{fig:yB}. The inset shows that $q_{y}^{*}/B^{*}$ depends linearly on the system size. From this observation, the amplitude of the NEFC seems to be proportional to the system size for sufficiently small magnetic fields.

Although above argument seems to be plausible, we remark that our system size is not very large to get asymptotic behavior, and hence it is fair to say that the system-size dependence of the NEFC still remains an open problem.

\section{\label{app:deltat}Dependence of temperature difference in the NEFC}
\begin{figure}[b]
\includegraphics[width=8.6cm]{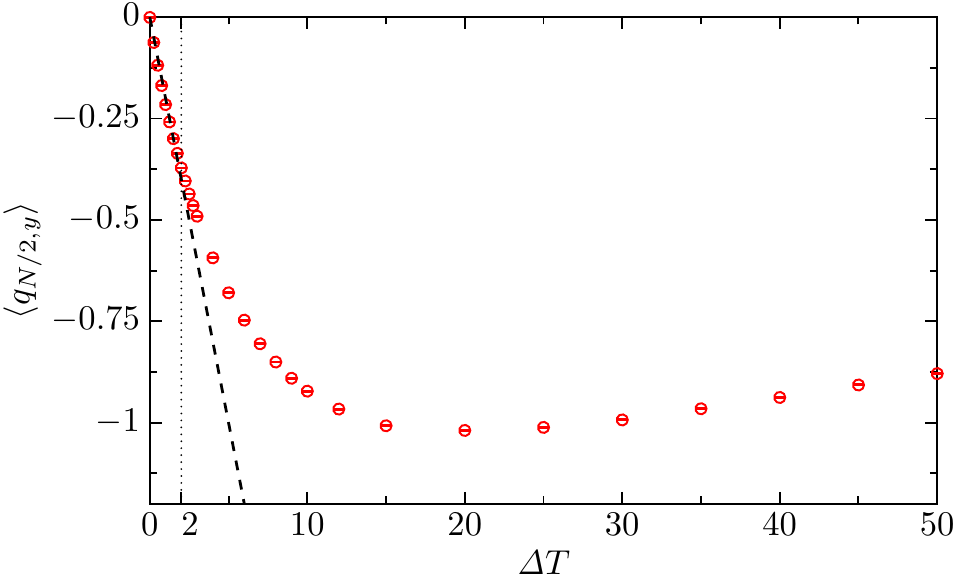}
\caption{\label{fig:yDT}Dependence of temperature difference in the transverse deviation at the central particle $\langle q_{N/2,y}\rangle$ for $eB=0.5$. The system size is $N=32$. The dashed line is a guiding line to highlight the linear dependence $\langle q_{N/2,y}\rangle\propto\varDelta T$ and the dotted lined represents $\varDelta T=2$. }
\end{figure}

We present the effect of temperature difference in the NEFC. We set the temperatures as $T_{\rm R}=1$ and $T_{\rm L}=T_{\rm R} +\varDelta T$ and consider the NEFC for many cases of $\varDelta T$.
We consider the system with $e^{2}/4\pi\varepsilon_{0}=0$ and show the $\varDelta T$ dependence of the deviation at the central particle $\langle q_{N/2 , y} \rangle$ for a fixed magnetic field $eB=0.5$. We show the result in Fig. \ref{fig:yDT}. As shown in the figure, up to $\varDelta T \sim 2$, the gradient is almost constant. This implies that this regime is regarded as the linear response regime. Beyond the linear response regime, i.e., for $\varDelta T>2$, the absolute value of the gradient begins to decrease.

\section{\label{app:linear-response}LINEAR RESPONSE FORMULA}

In this appendix, we derive the linear response formula~(\ref{eq:L00})-(\ref{eq:Lij}). We first consider the linear response for a static force $f_{i}$ and derive Eqs.~(\ref{eq:L0i}) and~(\ref{eq:Lij}). Then, we consider the linear response for a temperature difference $\varDelta T$ and derive Eqs.~(\ref{eq:L00}) and~(\ref{eq:Li0}).

\subsection{\label{sapp:linear-response-formula-f}Linear response for a static force in the $y$-direction}

We set the temperatures to $T_{\text{L}}=T_{\text{R}}=T$ and apply a static force $f_{i}$ to the $i$th particle in the $y$-direction. At equilibrium, the distribution of the phase-space variables $\bm{\varGamma}=(\{\bm{q}_{i}\}_{i=1}^{N},\{\bm{v}_{i}\}_{i=1}^{N})$ is expressed by the canonical distribution:
\begin{eqnarray}
P_{f_{i}}(\bm{\varGamma})
&=&\frac{e^{-[H(\bm{\varGamma})-f_{i}q_{i,y}]/ T}}{Z_{f_{i}}},
\label{a:eq:Pf}\\
Z_{f_{i}}
&:=&\int d\bm{\varGamma}\,e^{-\left[H(\bm{\varGamma})-f_{i}q_{i,y}\right]/ T}.
\label{a:eq:Zf}
\end{eqnarray}
We consider a sufficiently small $f_{i}$. Then, the partition function is approximated as $Z_{f_{i}}\approx Z(1+f_{i}\langle q_{i,y}\rangle_{\text{eq}})
=Z$, where $Z$ is defined in Eq.~(\ref{eq:canonical-distribution}). Then, Eq.~(\ref{a:eq:Pf}) is approximated as
\begin{eqnarray}
P_{f_{i}}(\bm{\varGamma})&\approx&P_{\text{eq}}(\bm{\varGamma})+\frac{f_{i}}{ T}q_{i,y}P_{\text{eq}}(\bm{\varGamma}),
\label{a:eq:Pf-approx}
\end{eqnarray}
where $P_{\text{eq}}(\bm{\varGamma})$ is defined in Eq.~(\ref{eq:canonical-distribution}). Therefore, the linear response of an arbitrary quantity $A(\bm{\varGamma})$ is expressed as
\begin{eqnarray}
\langle\delta A\rangle_{f_{i}}&:=&\langle A\rangle_{f_{i}}-\langle A\rangle_{\text{eq}}
=\frac{f_{i}}{ T}\langle Aq_{i,y}\rangle_{\text{eq}}.
\label{a:eq:A-f}
\end{eqnarray}
Here, the symbol $\langle\ \cdot\ \rangle_{f_{i}}$ represents the average with respect to Eq.~(\ref{a:eq:Pf}).

The linear response of the heat current $J$ is obtained as $\langle J\rangle_{f_{i}}=f_{i}\langle Jq_{i,y}\rangle_{\text{eq}}/ T=0$. The corresponding linear-response coefficient is
\begin{eqnarray}
L_{0i}:=\frac{\langle J\rangle_{f_{i}}}{f_{i}}=0.
\label{a:eq:L0i}
\end{eqnarray}

The linear response of the transverse deviation is expressed as $\langle q_{i,y}\rangle_{f_{j}}=f_{j}\langle q_{i,y}q_{j,y}\rangle_{\text{eq}}/ T$. Then, the linear-response coefficient is obtained as
\begin{eqnarray}
L_{ij}:=\frac{\langle q_{i,y}\rangle_{f_{j}}}{f_{j}}=\frac{1}{ T}\langle q_{i,y}q_{j,y}\rangle_{\text{eq}}.
\label{a:eq:Lij}
\end{eqnarray}

\subsection{\label{sapp:linear-response-formula-delT}Linear response formula for temperature difference}

Here, we derive Eqs.~(\ref{eq:L00}) and~(\ref{eq:Li0}) following Ref.~\cite{kundu-dhar-narayan-2009}. We set the temperature of the heat baths to $T_{\text{L}}=T+\varDelta T/2$ and $T_{\text{R}}=T-\varDelta T/2$. In this situation, the time-evolution of the distribution function $P(\bm{\varGamma},t)$ is expressed by the following Fokker-Planck equation:
\begin{eqnarray}
\frac{\partial}{\partial t}P(\bm{\varGamma},t)&=&(\mathbb{L}^{B}+\delta\mathbb{L})P(\bm{\varGamma},t).
\label{a:eq:FP-eq}
\end{eqnarray}
Here, the operators $\mathbb{L}^{B}=\mathbb{L}_{\text{H}}^{B}+\mathbb{L}_{\text{b}}$ and $\delta \mathbb{L}$ are defined by
\begin{widetext}
\begin{eqnarray}
\mathbb{L}_{\text{H}}^{B}&:=&\sum_{i=1}^{N}\left[\sum_{\alpha=x,y}\left(-\frac{\partial}{\partial q_{i,\alpha}}v_{i,\alpha}-\frac{\partial}{\partial v_{i,\alpha}}\frac{F_{i+1,i,\alpha}-F_{i,i-1,\alpha}}{m}\right)
-\frac{eB}{m}\left(\frac{\partial}{\partial v_{i,x}}v_{i,y}-\frac{\partial}{\partial v_{i,y}}v_{i,x}\right)\right],
\label{a:eq:def-LH}
\end{eqnarray}
\end{widetext}
\begin{eqnarray}
\mathbb{L}_{\text{b}}
&\!:=\!&
\frac{\gamma}{m}\sum_{i=1\,{\rm and}\,N}\sum_{\alpha=x,y}
\frac{\partial}{\partial v_{i,\alpha}}
\left(v_{i,\alpha}+\frac{ T}{m}\frac{\partial}{\partial v_{i,\alpha}}\right)\!,~~~~~
\label{a:eq:def-Lb}\\
\delta\mathbb{L}
&\!:=\!&
\frac{\gamma  \varDelta T}{2m^{2}}
\sum_{i=1}^{N}\sum_{\alpha=x,y}(\delta_{i1}-\delta_{iN})\frac{\partial^{2}}{\partial v_{i,\alpha}^{2}}.
\label{a:eq:def-dL}
\end{eqnarray}
Note that $\mathbb{L}^{B}$ is independent of $\varDelta T$. The canonical distribution Eq.~(\ref{eq:canonical-distribution}) satisfies $\mathbb{L}^{B}P_{\text{eq}}(\bm{\varGamma})=0$ and the detailed balance relation as an operator identity:
\begin{eqnarray}
\mathbb{L}^{B}P_{\text{eq}}(\bm{\varGamma})=P_{\text{eq}}(\bm{\varGamma})\vartheta\mathbb{L}^{\dag-B}.
\label{a:eq:detailed-balance}
\end{eqnarray}
The operator $\mathbb{L}^{\dag B}$ is the adjoint of $\mathbb{L}^{B}$ and written as
\begin{eqnarray}
\mathbb{L}^{\dag B}
&\!\!=\!\!&
-\mathbb{L}_{\text{H}}^{B}+\mathbb{L}_{\text{b}}^{\dag},
\label{a:eq:def-L-dag}\\
\mathbb{L}_{\text{b}}^{\dag}
&\!\!=\!\!&
\frac{\gamma}{m}\!\sum_{i=1\,{\rm and}\,N}\sum_{\alpha=x,y}\!
\left(\!
-v_{i,\alpha}
\!+\!
\frac{ T}{m}\frac{\partial}{\partial v_{i,\alpha}}
\right)\!
\frac{\partial}{\partial v_{i,\alpha}}\!.~~~~~~
\label{a:eq:def-Lb-dag}
\end{eqnarray}
The symbol $\vartheta$ in Eq.~(\ref{a:eq:detailed-balance}) acts on an operator and changes the sign of all $\bm{v}_{i}$.

Suppose that the temperature difference is set to $\varDelta T=0$ for $t<0$, and the distribution is the canonical distribution (\ref{eq:canonical-distribution}). Then, a finite temperature difference $\varDelta$ is switched on at $t=0$. For $t\ge0$, the formal solution of (\ref{a:eq:FP-eq}) is expressed as
\begin{eqnarray}
P^{B}(\bm{\varGamma},t)&=&e^{(\mathbb{L}^{B}+\delta\mathbb{L})t}P_{\text{eq}}(\bm{\varGamma})\nonumber\\
&\approx&P_{\text{eq}}(\bm{\varGamma})+\int_{0}^{t} ds\,e^{\mathbb{L}^{B}s}\delta\mathbb{L}P_{\text{eq}}(\bm{\varGamma}).
\label{a:eq:P-approx}
\end{eqnarray}
Here, we neglect the higher order terms of $\varDelta T$. Then, the linear response of an arbitrary quantity $A(\bm{\varGamma})$ is expressed as 
\begin{eqnarray}
\langle\delta A\rangle^{B}_{\varDelta T}&:=&\langle A\rangle_{\varDelta T}^{B}-\langle A\rangle_{\text{eq}}\nonumber\\
&=&\int_{0}^{\infty} dt\int d\bm{\varGamma}\,A(\bm{\varGamma})\, e^{\mathbb{L}^{B}t}\delta\mathbb{L}P_{\text{eq}}(\bm{\varGamma}).~~~
\label{a:eq:delA}
\end{eqnarray}
Here, the symbol $\langle\ \cdot\ \rangle_{\varDelta T}^{B}$ represents the average with respect to the steady state distribution.

One can calculate $\delta \mathbb{L} P_{\text{eq}}(\bm{\varGamma})$ as follows:
\begin{eqnarray}
\delta\mathbb{L}P_{\text{eq}}(\bm{\varGamma})
&\!\!=\!\!&
\frac{\varDelta T}{ T^{2}}I(\bm{\varGamma})P_{\text{eq}}(\bm{\varGamma}).\label{a:eq:dL-P}\\
I(\bm{\varGamma})
&\!\!:=\!\!&
\frac{\gamma}{m}
\!\!\left[
\left(\frac{m\bm{v}_{1}^{2}}{2}\!-\! T\!\right)
\!-\!
\left(\frac{m\bm{v}_{N}^{2}}{2}\!-\! T\!\right)
\right]\!\!.~~~~~~
\label{a:eq:def-I}
\end{eqnarray}
Note that the function $I(\bm{\varGamma})$ is related to the heat current at the boundary~(\ref{eq:jL}) and (\ref{eq:jR}). Denote the net heat current measured at the boundary by $J_{\text{b}}:=\left(j_{\text{L}}-j_{\text{R}}\right)/2$. Then, it is demonstrated by straightforward calculation that $I(\bm{\varGamma})$ is related to the noise average of $J_{\text{b}}$,
\begin{eqnarray}
I(\bm{\varGamma}(t))=-\bm{\langle}\!\langle J_{\text{b}}(t)\rangle\!\bm{\rangle}.
\label{a:eq:IJ}
\end{eqnarray}

Using the detailed balance relation (\ref{a:eq:detailed-balance}), we can further calculate as
\begin{eqnarray}
\langle \delta A\rangle^{B}_{\varDelta T}
&\!=\!&\frac{\varDelta T}{ T^{2}}
\int_{0}^{\infty}\hspace{-6pt} dt\!\!
\int\hspace{-3pt} d\bm{\varGamma}
A(\bm{\varGamma})
 e^{\mathbb{L}^{B}t}\left[I(\bm{\varGamma})P_{\text{eq}}(\bm{\varGamma})\right],\nonumber\\
&\!=\!&\frac{\varDelta T}{ T^{2}}
\int_{0}^{\infty}\hspace{-6pt} dt\!\!
\int\hspace{-3pt} d\bm{\varGamma}
A(\bm{\varGamma})
P_{\text{eq}}(\bm{\varGamma})
 e^{\vartheta\mathbb{L}^{\dag-B}t}I(\bm{\varGamma}),\nonumber\\
&\!=\!&\frac{\epsilon_{A}\varDelta T}{ T^{2}}
\int_{0}^{\infty}\hspace{-6pt} dt\!\!
\int\hspace{-3pt} d\bm{\varGamma}
P_{\text{eq}}(\bm{\varGamma})
A(\bm{\varGamma})
 e^{\mathbb{L}^{\dag-B}t}I(\bm{\varGamma}),\nonumber\\
&\!=\!&\frac{\epsilon_{A}\varDelta T}{ T^{2}}
\int_{0}^{\infty}\hspace{-6pt} dt\!\!
\int\hspace{-3pt} d\bm{\varGamma}
P_{\text{eq}}(\bm{\varGamma})
A(\bm{\varGamma})
I(\bm{\varGamma}^{-B}(t)).~~~~~~~
\label{a:eq:response}
\end{eqnarray}
Here, the variable $\epsilon_{A}$ assumes the value $\epsilon_{A}=1$ if $A$ is an even function of the velocities and $\epsilon_{A}=-1$ if $A$ is an odd function of the velocities. From the second line to the third line, the variables $\bm{v}_{i}$ are changed into $-\bm{v}_{i}$. The variable $\bm{\varGamma}^{B}(t)$ represents the positions and velocities at time $t$, which evolves under a magnetic field $B$ from the initial state $\bm{\varGamma}$.

As the $A(\bm{\varGamma})$ in Eq.~(\ref{a:eq:response}) is uncorrelated with the noises arising during $t\ge 0$, it holds that $\bm{\langle}\!\langle A(\bm{\varGamma})J_{\text{b}}(t)\rangle\!\bm{\rangle}=A(\bm{\varGamma})\bm{\langle}\!\langle J_{\text{b}}(t)\rangle\!\bm{\rangle}$. Combining this with Eq.~(\ref{a:eq:IJ}), we can obtain
\begin{eqnarray}
A(\bm{\varGamma})I(\bm{\varGamma}(t))
=-\bm{\langle}\!\langle A(\bm{\varGamma})J_{\text{b}}(t)\rangle\!\bm{\rangle}.
\label{a:eq:AI-AJ}
\end{eqnarray}
Substituting Eq.~(\ref{a:eq:AI-AJ}) into Eq.~(\ref{a:eq:response}), we obtain
\begin{eqnarray}
\langle \delta A\rangle^{B}_{\varDelta T}=-\epsilon_{A}\frac{\varDelta T}{ T^{2}}\int_{0}^{\infty} dt\,\langle A(0)J_{\text{b}}(t)\rangle_{\text{eq}}^{-B}.
\label{a:eq:LR-delT-A}
\end{eqnarray}

Using the continuity equation of the local energy, it is demonstrated that the boundary current $J_{\text{b}}(t)$ in Eq.~(\ref{a:eq:LR-delT-A}) can be replaced by the bulk current $J(t)$, if the quantity $A$ satisfies the following two conditions (see Ref.~\cite{kundu-dhar-narayan-2009} for more details)
\begin{eqnarray}
\langle A\rangle_{\rm eq}=0,
\quad\text{and}\quad
\langle A\varepsilon_{i}\rangle_{\text{eq}}=0\quad\text{for all}\ i.
\label{a:eq:conds}
\end{eqnarray}
Here, $\varepsilon_{i}$ is the local energy defined by $\varepsilon_{i}:=m|\bm{v}_{i}|^{2}/2+V(|\bm{r}_{i+1,i}|)$ . The heat current $J$ and deviation $q_{i,y}$ satisfy Eq.~(\ref{a:eq:conds}). Furthermore, using the detailed balance relation~(\ref{a:eq:detailed-balance}), it is shown that
\begin{eqnarray}
\langle A(0)J(t)\rangle_{\text{eq}}^{-B}=-\epsilon_{A}\langle A(t)J(0)\rangle_{\text{eq}}^{B}.
\label{a:eq:AJ}
\end{eqnarray}
From these considerations, the linear response of the heat current and transverse deviation are respectively expressed as
\begin{eqnarray}
\langle J\rangle^{B}_{\varDelta T}
&=&\frac{\varDelta T}{ T^{2}}\int_{0}^{\infty} dt\,\langle J(t)J(0)\rangle_{\text{eq}}^{B},
\label{a:eq:LR-delT-J-3}\\
\langle q_{i,y}\rangle^{B}_{\varDelta T}&=&\frac{\varDelta T}{ T^{2}}\int_{0}^{\infty} dt\,\langle q_{i,y}(t)J(0)\rangle_{\text{eq}}^{B}.
\label{a:eq:LR-delT-y-3}
\end{eqnarray}
The corresponding linear-response coefficients are obtained as
\begin{eqnarray}
L_{00}(B)
&\!\!:=\!\!&
\frac{\langle J\rangle_{\varDelta T}^{B}}{\varDelta T}=\frac{1}{ T^{2}}\int_{0}^{\infty}\hspace{-4pt} dt\,\langle J(t)J(0)\rangle_{\text{eq}}^{B},~~~~~~
\label{a:eq:L00}\\
L_{i0}(B)
&\!\!:=\!\!&\frac{\langle q_{i,y}\rangle_{\varDelta T}^{B}}{\varDelta T}=\frac{1}{ T^{2}}\int_{0}^{\infty}\hspace{-4pt} dt\,\langle q_{i,y}(t)J(0)\rangle_{\text{eq}}^{B}.~~~~~~
\label{a:eq:Li0}
\end{eqnarray}
Note that for the element $L_{00}(B)$, Eq.~(\ref{a:eq:AJ}) derives the Onsager-Casimir symmetry $L_{00}(-B)=L_{00}(B)$.

\section{\label{a:sec:harmonic-chain}EXACT EXPRESSION OF NET HEAT IN THE CASE OF LINEAR FORCES}

In this appendix, we analytically show that the inverse effect cannot occur in the case of linear dynamics, where the natural length is set to zero. Suppose the external force $f(t)$ in Eq.~(\ref{eq:external-force}) is applied to the $i_{0}$th particle in the $y$-direction. The equations of motion are expressed as
\begin{eqnarray}
m\dot{v}_{i,x}&=&k(q_{i+1,x}+q_{i-1,x}-2q_{i,x})+eBv_{i,y}\nonumber\\
&+&\delta_{i,1}(-\gamma v_{i,x}+\eta_{\text{L},x})
+\delta_{i,N}(-\gamma v_{i,x}+\eta_{\text{R},x}),~~~~
\label{a:eq:EOM-x-harmonic}\\
m\dot{v}_{i,y}&=&k(q_{i+1,y}+q_{i-1,y}-2q_{i,y})-eBv_{i,x}+\delta_{i,i_{0}}f(t)\nonumber\\
&+&\delta_{i,1}(-\gamma v_{i,y}+\eta_{\text{L},y})
+\delta_{i,N}(-\gamma v_{i,y}+\eta_{\text{R},y}).~~~~
\label{a:eq:EOM-y-harmonic}
\end{eqnarray}

Here, we solve the equations of motion by the method of the Green's function~\cite{dhar-saito-2016}. For convenience, we use the following vector notation:
\begin{eqnarray}
\bm{Q}^{T}&=&[q_{1,x},\cdots,q_{N,x},q_{1,y},\cdots,q_{N,y}],
\label{a:eq:Q}\\
\bm{V}^{T}&=&[v_{1,x},\cdots,v_{N,x},v_{1,y},\cdots,v_{N,y}],
\label{a:eq:V}
\end{eqnarray}
where the superscript $^{T}$ stands for the transpose of a vector or matrix. The noises and the external force are also denoted by vectors
\begin{eqnarray}
[\bm{\eta}_{\text{L}}(t)]_{i}&=&\delta_{i,1}\eta_{\text{L},x}(t)+\delta_{i,N+1}\eta_{\text{L},y}(t),
\label{a:eq:etaL}\\
\protect[\bm{\eta}_{\text{R}}(t)]_{i}&=&\delta_{i,N}\eta_{\text{R},x}(t)+\delta_{i,2N}\eta_{\text{R},y}(t),
\label{a:eq:etaR}\\
\protect[\bm{f}(t)]_{i}&=&\delta_{i,N+i_{0}}f(t).
\label{a:eq:f}
\end{eqnarray}
In addition, we introduce $2N\times 2N$ matrices:
\begin{eqnarray}
\bar{\mathsf{M}}&=&\begin{bmatrix}
\mathsf{M}&\mathsf{O}\\\mathsf{O}&\mathsf{M}
\end{bmatrix},\quad
\bar{\mathsf{K}}=\begin{bmatrix}
\mathsf{K}&\mathsf{O}\\\mathsf{O}&\mathsf{K}
\end{bmatrix},
\label{a:eq:M-K}\\
\bar{\mathsf{B}}&=&\begin{bmatrix}
\mathsf{O}&\mathsf{B}\\\mathsf{-B}&\mathsf{O}
\end{bmatrix},\quad
\bar{\mathsf{R}}_{\text{L,R}}=\begin{bmatrix}
\mathsf{R}_{\text{L,R}}&\mathsf{O}\\
\mathsf{O}&\mathsf{R}_{\text{L,R}}
\end{bmatrix},
\label{a:eq:B-R}
\end{eqnarray}
with $\mathsf{M}_{i,j}:=m\delta_{i,j}$, $\mathsf{K}_{i,j}:=k(-\delta_{i+1,j}-\delta_{i-1,j}+2\delta_{i,j})$, $\mathsf{B}_{i,j}:=eB\delta_{i,j}$, $(\mathsf{R}_{\text{L}})_{i,j}:=\gamma\delta_{i,1}\delta_{i,j}$, and $(\mathsf{R}_{\text{R}})_{i,j}:=\gamma\delta_{i,N}\delta_{i,j}$. Using these notations, the equations of motion are expressed as
\begin{eqnarray}
\bar{\mathsf{M}}\dot{\bm{V}}=-\bar{\mathsf{K}}\bm{Q}+\bar{\mathsf{B}}\bm{V}-\bar{\mathsf{R}}_{\text{L}}\bm{V}-\bar{\mathsf{R}}_{\text{R}}\bm{V}+\bm{\eta}_{\ell}+\bm{\eta}_{r}+\bm{f},~~~
\label{a:eq:EOM}
\end{eqnarray}

To solve Eq.~(\ref{a:eq:EOM}), we introduce the Fourier transforms:
\begin{eqnarray}
\tilde{\bm{Q}}(\omega)&\!=\!\!&
\int_{-\infty}^{\infty}\!\!\!\! dt\,
\bm{Q}(t) e^{\text{i}\omega t},\ \,
\tilde{\bm{V}}(\omega)\!=\!\!
\int_{-\infty}^{\infty}\!\!\!\! dt\,
\bm{V}(t) e^{\text{i}\omega t},~~~~~~
\label{a:eq:F-tr-X-V}\\
\tilde{\bm{\eta}}_{\text{L,R}}(\omega)&\!=\!\!&
\int_{-\infty}^{\infty}\!\!\!\! dt\,
\bm{\eta}_{\text{L,R}}(t) e^{\text{i}\omega t},\ \,
\tilde{\bm{f}}(\omega)\!=\!\!
\int_{-\infty}^{\infty}\!\!\!\! dt\,
\bm{f}(t) e^{\text{i}\omega t}.~~~~~~
\label{a:eq:F-tr-eta-f}
\end{eqnarray}
Then, the solution of Eq.~(\ref{a:eq:EOM}) is expressed as
\begin{eqnarray}
\tilde{\bm{Q}}(\omega)&=&\bar{\mathsf{G}}_{B}^{+}(\omega)\left[\tilde{\bm{\eta}}_{\text{L}}(\omega)+\tilde{\bm{\eta}}_{\text{R}}(\omega)+\tilde{\bm{f}}(\omega)\right],
\label{a:eq:solution-Q}\\
\tilde{\bm{V}}(\omega)&=&-{\rm i}\omega\bar{\sf G}_{B}^{+}(\omega)\left[\tilde{\bm{\eta}}_{\text{L}}(\omega)+\tilde{\bm{\eta}}_{\text{R}}(\omega)+\tilde{\bm{f}}(\omega)\right].~~~
\label{a:eq:solution-V}
\end{eqnarray}
Here, the $2N\times 2N$ matrix $\bar{\mathsf{G}}_{B}^{+}(\omega)$ is the Green's function defined as 
\begin{eqnarray}
\bar{\mathsf{G}}_{B}^{+}(\omega):=\left[-\omega^{2}\bar{\mathsf{M}}+\bar{\mathsf{K}}+\text{i}\omega\bar{\mathsf{B}}-\text{i}\omega\bar{\mathsf{R}}_{\text{L}}-\text{i}\omega\bar{\mathsf{R}}_{\text{R}}\right]^{-1}.~~~~~
\label{a:eq:def-G}
\end{eqnarray}
Note that $\bar{\mathsf{G}}_{B}^{+}(\omega)$ is simplified as
\begin{eqnarray}
\bar{\mathsf{G}}_{B}^{+}(\omega)=\begin{bmatrix}
\mathsf{F}_{B}^{+}(\omega)&-\text{i}\omega\mathsf{F}_{B}^{+}(\omega)\mathsf{B}\mathsf{G}^{+}(\omega)\\
\text{i}\omega\mathsf{F}_{B}^{+}(\omega)\mathsf{B}\mathsf{G}^{+}(\omega)&\mathsf{F}_{B}^{+}(\omega)
\end{bmatrix},~~~~~~
\label{a:eq:G}
\end{eqnarray}
with
\begin{eqnarray}
\mathsf{G}^{+}(\omega)&:=&\left[-\omega^{2}\mathsf{M}+\mathsf{K}-\text{i}\omega\mathsf{R}_{\text{L}}-\text{i}\omega\mathsf{R}_{\text{R}}\right]^{-1},
\label{a:eq:def-G-mat}\\
\mathsf{F}_{B}^{+}(\omega)&:=&\left[\left[\mathsf{G}^{+}(\omega)\right]^{-1}-\omega^{2}\mathsf{B}\mathsf{G}^{+}(\omega)\mathsf{B}\right]^{-1}.
\label{a:eq:def-F-mat}
\end{eqnarray}

Using the solution~(\ref{a:eq:solution-V}), we calculate the heat $\mathcal{Q}_{\text{L,R}}$ flowing from the system into the heat bath during the period $\tau$, which is defined in Eq.~(\ref{eq:def-Q}). The heat current at the boundary, defined in Eqs.~(\ref{eq:jL}) and~(\ref{eq:jR}), is expressed as
\begin{eqnarray}
j_{\mu}=\text{Tr}\left(-\bar{\mathsf{R}}_{\mu}\bm{V}\bm{V}^{T}+\bm{\eta}_{\mu}\bm{V}^{T}\right)
\ \,\text{with}\ \,\mu=\text{L, R}.~~~
\label{a:eq:jL-jR}
\end{eqnarray}
The calculation is lengthy albeit straightforward. We use the relation
\begin{eqnarray}
\bm{\langle}\!\langle\tilde{\bm{\eta}}_{\mu}(\omega)\tilde{\bm{\eta}}_{\nu}^{T}(\omega')\rangle\!\bm{\rangle}=4\pi  T\delta_{\mu\nu}\delta(\omega+\omega')\bar{\mathsf{R}}_{\mu},
\label{a:eq:FDR}
\end{eqnarray}
and the specific form of the external force~(\ref{eq:external-force}). The result is expressed as
\begin{widetext}
\begin{eqnarray}
\langle \mathcal{Q}_{\text{L}}(\tau)\rangle&=&
\frac{\gamma f_{0}^{2}}{2}\tau\Omega^{2}\biggl\{
[\mathsf{F}_{B}^{+}(2\Omega)]_{1,i_{0}}[\mathsf{F}_{B}^{-}(2\Omega)]_{1,i_{0}}
+4B^{2}[\mathsf{F}^{+}_{B}(2\Omega)\mathsf{G}^{+}(2\Omega)]_{1,i_{0}}[\mathsf{F}^{-}_{B}(2\Omega)\mathsf{G}^{-}(2\Omega)]_{1,i_{0}}\biggr\},
\label{a:eq:QL}\\
\langle \mathcal{Q}_{\text{R}}(\tau)\rangle&=&\frac{\gamma f_{0}^{2}}{2}\tau\Omega^{2}\biggl\{
[\mathsf{F}_{B}^{+}(2\Omega)]_{N,i_{0}}[\mathsf{F}_{B}^{-}(2\Omega)]_{N,i_{0}}+4B^{2}[\mathsf{F}^{+}_{B}(2\Omega)\mathsf{G}^{+}(2\Omega)]_{N,i_{0}}[\mathsf{F}^{-}_{B}(2\Omega)\mathsf{G}^{-}(2\Omega)]_{N,i_{0}}\biggr\}.
\label{a:eq:QR}
\end{eqnarray}
\end{widetext}
Here, $\Omega:=2\pi/\tau$ is the angular frequency of the external force. The matrices $\mathsf{G}^{-}(\omega)$ and $\mathsf{F}^{-}_{B}(\omega)$ are the Hermitian conjugate of $\mathsf{G}^{+}(\omega)$ and $\mathsf{F}^{+}_{B}(\omega)$.

The results~(\ref{a:eq:QL}) and~(\ref{a:eq:QR}) are invariant to the magnetic field reversal $B\leftrightarrow -B$ because of $\mathsf{F}^{\pm}_{-B}(\omega)=\mathsf{F}^{\pm}_{B}(\omega)$. Furthermore, the matrices $\mathsf{G}^{\pm}(\omega)$ and $\mathsf{F}^{\pm}_{B}(\omega)$ have the symmetric property of $[\mathsf{G}^{\pm}(\omega)]_{i,j}=[\mathsf{G}^{\pm}(\omega)]_{N+1-j,N+1-i}$ and $[\mathsf{F}^{\pm}_{B}(\omega)]_{i,j}=[\mathsf{F}^{\pm}_{B}(\omega)]_{N+1-j,N+1-i}$. Therefore, the heat as a function of $i_{0}$ has the symmetry,
\begin{eqnarray}
\langle \mathcal{Q}_{\text{L}}\rangle(i_{0})=\langle \mathcal{Q}_{\text{R}}\rangle(N+1-i_{0}).
\label{a:eq:symmetric-relation-Q}
\end{eqnarray}
This implies that $\langle \mathcal{Q}_{\text{L}}-\mathcal{Q}_{\text{R}}\rangle=0$ if the external forces are applied to the particles located symmetrically with respect to the center of the system, which is the present setup in the main text. Thus, the inverse effect is prohibited.

\bibliography{apssamp}

\begin{thebibliography}{99}
\bibitem{tltr}
This temperature regime is relatively low in our parameter set. For the case with the parameter $e^{2}/4\pi \epsilon_0 =10$ in the equilibrium situation with the temperature $T=1.5$, the equilibrium standard deviation of stretch in the bulk $ \sqrt{\langle |{\bm r}|^{2}\rangle_{\rm eq}-\langle|{\bm r}|\rangle_{\rm eq}^{2}}  \sim 1.2$, which is smaller than the natural length $\ell=10$. This parameter set is advantageous in accurate numerical calculation, since each particle does not get close for the finite Coulomb interaction.
\bibitem{fn-2}
The left-right reversal is expressed as $q_{i,x}\to(N+1)\ell-q_{N+1-i,x}$. This is equivalent to the magnetic field reversal $B\to-B$. Hence, in the case of zero magnetic field, the dynamics is invariant under the left-right reversal. Since the sign of the net heat $\mathcal{Q}_{\text{L}}-\mathcal{Q}_{\text{R}}$ (defined in Eq.(\ref{eq:def-Q})) is reversed under the left-right reversal, the net heat is always zero in the absence of magnetic fields.
\end{thebibliography}

\end{document}